# Model-free Anomaly Detection for Dynamical Systems with Gaussian Processes

**Alejandro Penacho Riveiros** * **Nicola Bastianello** *
**Matthieu Barreau** *

* *Digital Futures and KTH Royal Institute of Technology, Stockholm, Sweden (email: alejpr@kth.se, nicolba@kth.se, barreau@kth.se)*

**Abstract:** In this paper we address the problem of detecting differences or anomalies in a dynamical system, based on historical data of nominal operations. This problem encompasses quality control, where newly manufactured systems are tested against desired nominal operations, and the detection of changes in the dynamics due to degradation or repairs. We propose a model-free approach based on Gaussian processes (GPs). The idea is to train offline a GP based on nominal data, which is then deployed online to detect whether measurements of the system's state are compatible with nominal operations or if they deviate. Detecting this deviation is made more challenging by the presence of process and measurement noise, which might obfuscate deviations in the dynamics. The detection then is based on a threshold that ensures a specific false positive rate. We showcase the promising performance of the proposed method with two systems, and highlight several interesting future research questions.

## 1. INTRODUCTION

Consider a classical problem in quality control: first, we have a set of products considered satisfactory. We then receive a new one, and we must determine if, by its properties, it belongs to this initial set. This is the *discriminant analysis* problem, which attracted considerable attention in the field of statistics during the first half of the 20[th], leading to the development of tools like the $\chi^2$-test or the Fisher-Irwin test. These tools are still used to this day in quality control and clinical testing, as discussed in Salsburg (1992).

The techniques developed during this period had a strong theoretical foundation, but were limited to static objects, in which all properties of interest could be observed at once. The extension to dynamical system took place mostly during the 70's and 80's, and was later synthesized in works such as Frank (1990) or Patton & Chen (1997), always with the focus of detecting anomalous or faulty behavior of a system. Although well grounded, these methods relied on a nominal prescribed model against which new data was to be compared. Unfortunately, such nominal model can not always be obtained, as some systems are too complex to be modeled accurately, or their properties evolve in unknown ways during their operation.

The work of was partially supported by the European Union's Horizon Research and Innovation Actions programme under grant agoreement No. 101070162, and by the Wallenberg AI, Autonomous Systems and Software Program (WASP) funded by the Knut and Alice Wallenberg Foundation.

When the nominal behavior of the system is only known through data, the usual approach is to first do system identification to obtain a model, and then apply model-based techniques (Simani et al. (2003)). There are, however, two limitations to this technique. First, the identification of a model from data always leads to some level of uncertainty which can be unfeasible to quantify, and must be taken into account to perform a correct analysis. Second, the models used for identification are parametric, so their expressiveness is limited, and their design requires some prior knowledge of the system under study.

The study of uncertainty quantification in system identification has received a good amount of attention, as covered by Latifi & Scoglio (2025), and have found some uses in fault and anomaly detection, like in Liao et al. (2024).

For non-parametric models, their development can be linked to the field of machine learning, which led to development of highly expressive architectures such as neural networks. These architectures are able to exploit the exponentialy increasing availability of computation resources to approximate functions of any kind, proved by Cybenko (1989). This has led to a major influx of data-driven fault detection techniques, capable of dealing with very complex systems for which no model is available: the paper by Neupane et al. (2025) offers a good review on the topic. Unfortunately, although effective in their purpose, these methods usually provide a yes/no diagnostic, without much consideration for statistical concerns.

Combining both advances, uncertainty quantification and non-parametric models, has proven challenging. Some approaches, like Bayesian Neural Networks, have been applied succesfully to fault detection, as in Maged & Xie

(2022). However, these techniques are very complex and usually require large quantities of data to work correctly.

A more satisfactory solution for non-parametric architectures with uncertainty quantification can be found in Gaussian processes, explained in detail by Rasmussen & Williams (2005). GPs have been applied succesfully to a wide variety of tasks, from classification (Opper & Winther (2000)) to control of dynamical systems (Scampicchio et al. (2025)). For the latter, researchers have been able to infuse GPs with a wide variety of properties: Beckers et al. (2022) imposed a Port-Hamiltonian structure to the resulting model, while Marino & Cicirello (2023) allowed for switching behavior of the resulting system. This makes them extremely versatile, as an user can select how expressive their model needs to be, from universal approximator to linear system, all under the same framework.

In this paper, we use GP to solve the following problem: given a "nominal" system which we know only through several trajectories generated by it, we must determine if new trajectory has been generated by this same model, or by a different one.

The motivation behind using GPs is their data efficiency and, most importantly, their capability to assess uncertainty in their approximations. The contributions of the paper are the following:

1. We use a GP to estimate the sampled-time dynamics of a system with the purpose of generating residuals for a new trajectory.
2. We fully characterize the distribution of the residuals produced by our method and use them to set a threshold adjusted to the rate of false positives allowed by the algorithm.
3. We discuss the trade-offs present in the detection of anomalous behavior, considering the relation between the difference in the dynamics between the nominal and non-nominal systems, the noise present in the system and the amount of data available for training the GP.

## 2. PROBLEM FORMULATION

In this section we introduce the model of the system considered in the paper, and then proceed with the problem statement.

### 2.1 System model

We consider a continuous autonomous system on which we have full state observations, with dynamics given by:

$$\begin{aligned} \dot{x}(t) &= f(x(t)) + \eta(t) \\ \hat{x}(t) &= x(t) + v(t) \end{aligned} \quad (1)$$

where $x, \hat{x}, \eta, v \in \mathbb{R}^{n_x}$, and $\eta(t)$ and $v(t)$ are continuous uncorrelated white noises. The system is sampled at a constant rate $\Delta T$:

$$\begin{aligned} x_{k+1} &= x((k+1)\Delta T) = x_k + F(x_k) + w_k \\ \hat{x}_k &= \hat{x}(k\Delta T) = x_k + v_k, \end{aligned} \quad (2)$$

In this paper, we consider a system to have a constant sampling time, as is the case with most systems. Under this condition, we can ignore the continuous dynamics presented in (1) and instead focus on the discrete dynamics. Since this approach is purely data-driven, we do not use a priori information about $f(x)$ and $F(x)$. We do, on the other hand, assume that the latter can be approximated by a GP.

### 2.2 System Discrimination Problem

The problem tackled in this paper is as follows: consider that we have a dataset $D$ of $N_D$ trajectories, each one of length $N_\ell$, produced by the system described in (2),

$$D = \left\{ \left\{ \hat{x}_k^\ell \right\}_{k=1}^{n_\ell} \right\}_{\ell=1}^{n_D}. \quad (3)$$

Note that the trajectories do not need to have the same lenght, but we will assume so for the sake simplicity. Although these trajectories may have different initial conditions, they have all been produced using the same transition function $F_D(x)$, and have known observation and model error covariances $\mathbb{E}\left[w_k w_k^T\right] = \Sigma_w$ and $\mathbb{E}\left[v_k v_k^T\right] = \Sigma_v$. From now on we use term *nominal system* to refer to a system whose sampled observations follow (2) with transition function $F_D(x)$, regardless of the covariance of $w$ and $v$, and we call a *nominal trajectory* any trajectory generated by the nominal system.

We then receive a new trajectory

$$Q = \{\hat{x}_k\}_{k=1}^{n_{\text{steps}}}, \quad (4)$$

produced by a system with known noise covariances $\Sigma_v$ and $\Sigma_w$ (which do not need to be the same as in $D$), but unknown transition function $F(x)$. The objective in this paper is to devise an algorithm that determines whether the trajectory was generated with transition function $F_D(x)$, known only through the dataset $D$, with a tunable level of confidence adjusted by the maximum acceptable false positive rate.

## 3. PROPOSED ALGORITHM

### 3.1 Algorithm description

To solve the problem defined in Section 2.2, we propose a technique in two stages. In the first stage, performed offline, we train a GP with the dataset $D$ to produce an estimation of the transition function $\hat{F}_D(x)$. In the second stage, done online, we compare the observed trajectory $Q$ with $\hat{F}_D(x)$ to obtain a residual $\varepsilon$ that quantifies its deviation from the nominal behavior. With our knowledge of the distribution of $\varepsilon$, we produce a $p$-value of $Q$ that can be used to determine whether it fits with the dataset $D$.

Note that the training of the GP must only be done once, and it can then be used to test as many trajectories $Q$ as needed.

### 3.2 Estimation of $\hat{F}_D$ with a Gaussian Process

The first step is to create a function $\hat{F}_D(x)$ that approximates the function $F_D(x)$, only from knowledge of the trajectories in $D$.

To do this, we first turn each trajectory in the dataset into a sequence of pairs of observations before and after every transition:

$$X_\ell = \begin{bmatrix} \hat{x}_1^\ell \\ \hat{x}_2^\ell \\ \vdots \\ \hat{x}_{N_\ell-1}^\ell \end{bmatrix}, \quad Y_\ell = \begin{bmatrix} \hat{x}_2^\ell \\ \hat{x}_3^\ell \\ \vdots \\ \hat{x}_{N_\ell}^\ell \end{bmatrix} \quad (5)$$

A complete dataset is obtained by concatenating all the trajectories:

$$X_D = \begin{bmatrix} X_1 \\ X_2 \\ \vdots \\ X_{N_D} \end{bmatrix}, \quad Y_D = \begin{bmatrix} Y_1 \\ Y_2 \\ \vdots \\ Y_{N_D} \end{bmatrix}, \quad (6)$$

where both $X_D$ and $Y_D$ belong to $\mathbb{R}^{n_D(n_\ell-1) \times n_x}$. The number of samples do not need to be too large: in our examples, we use around 20 trajectories with 10 points each. We define our GP estimation of $F_D$ as

$$\hat{F}_D(x) = k(x, X_D)[k(X_D, X_D)]^{-1} Y_D, \quad (7)$$

where $k$ is a kernel function, which must be positive semi-definite. Additionally, we can compute the covariance of the error of this approximation in two different points $x_1$ and $x_2$ by:

$$\mathbb{E}\left[\left(\hat{F}_D(x_1) - F(x_1)\right)\left(\hat{F}_D(x_2) - F(x_2)\right)^T\right] \\ = k(x_1, x_2) - k(x_1, X_D)[k(X_D, X_D)]^{-1} k(X_D, x_2) \quad (8)$$

The kernel determines how the GP estimates the function away from the training points, considerable discussion has taken place on how to choose the best for a particular problem. In our method we will use the squared-exponential function due to its popularity and its universal approximation property, as shown by Maged & Xie (2022). This kernel is defined as:

$$k(x, x') = \sigma_f^2 \exp[-(\|x - x'\|^2)/(2\ell^2)] + \delta_{x,x'}\sigma_n^2, \quad (9)$$

where the hyperparameters $\sigma_f$ and $\ell$ are optimized to maximize the log-likelihood of the data in $X_D$ and $Y_D$, and $\sigma_n^2$ is chosen to be equal to the diagonal elements of $\Sigma_w$.

### 3.3 Generation of residuals

Once the approximation $\hat{F}_D$ has been obtained, we can compute for each step a residual

$$\varepsilon_k = (\hat{x}_{k+1} - \hat{x}_k) - \hat{F}_D(\hat{x}_k) \quad (10)$$

that denotes how the observed transition of state is different from the one expected by the GP. These residuals can then be concatenated in a residual vector:

$$\varepsilon = \begin{bmatrix} \varepsilon_1 & \varepsilon_2 & \cdots & \varepsilon_{n_{\text{steps}}-1} \end{bmatrix}^T \in \mathbb{R}^{(n_{\text{steps}}-1) \cdot n_x}. \quad (11)$$

Intuitively, a large error is related to a difference between the dynamics of $Q$ and $F_D$, and a common approach in system discrimination is to set a threshold $\theta$ for the squared norm of the error, so that we declare that $Q$ was not generated by $F_D$ if $\|\varepsilon\|^2 > \theta$. However, designing this threshold is not trivial, as the residuals are affected by several components, most of them unrelated to actual differences between the transition functions. Instead, we will characterize the distribution of $\|\varepsilon\|^2$, so when we obtain the residual of a trajectory $Q$, we can determine how likely it is to be produced by $F_D(x)$.

### 3.4 Characterization of residuals

There are several sources of uncertainty in the computation of the residual. In order to determine its distribution, we must first decompose it into different components. In particular, we consider 4 components of the residual:

$$\varepsilon_k = (\hat{x}_{k+1} - \hat{x}_k) - \hat{f}_D(y_k) \\ = \varepsilon_{n,k} + \varepsilon_{F,k} + \varepsilon_{\text{GP},k} + \varepsilon_{\text{obs},k}, \quad (12)$$

with each component defined as follows:

$$\begin{aligned} \text{Noise error:} \quad & \varepsilon_{n,k} = (\hat{x}_{k+1} - \hat{x}_k) - F_Q(x_k) \\ \text{Dynamics error:} \quad & \varepsilon_{F,k} = F_Q(x_k) - F_D(x_k) \\ \text{GP error:} \quad & \varepsilon_{\text{GP},k} = F_D(x_k) - \hat{F}_D(x_k) \\ \text{Observation error:} \quad & \varepsilon_{\text{obs},k} = \hat{F}_D(x_k) - \hat{F}_D(\hat{x}_k) \end{aligned} \quad (13)$$

In these expressions, we define $F_Q$ as the transition function that has produced the trajectory $Q$, which might or might not be $F_D$. We can now explain each component and estimate their distribution.

*Noise error ($\varepsilon_{n,k}$)* The first component is the difference between the observed evolution of the state of the system and its actual evolution, due purely to noise in the observation and in the dynamics of the system. It can be characterized as:

$$\varepsilon_{n,k} = (\hat{x}_{k+1} - \hat{x}_k) - F_Q(x_k) \\ = (x_{k+1} + v_{k+1} - x_k - v_k) - (x_{k+1} - x_k - w_k) \quad (14) \\ = v_{k+1} - v_k + w_k$$

*Dynamics error ($\varepsilon_{F,k}$)* This error is due to differences between the transition function of the nominal system and the system that produced $Q$. Since the transition functions are deterministic, this component is deterministic as well, and depends only on the state $x_k$:

$$\varepsilon_{F,k} = F_Q(x_k) - F_D(x_k) = \varepsilon_F(x_k) \quad (15)$$

*GP error ($\varepsilon_{\text{GP},k}$)* The GP error is due to the difference between the estimation of the transition function $\hat{F}_D(x)$ and the actual transition function $F_D(x)$. As we chose a GP for the estimation, this error is expected to follow a normal distribution with zero mean and covariance function given by

$$\Sigma_{\text{GP}}(x_1, x_2) = \\ k(x_1, x_2) - k(x_1, X_D)[k(X_D, X_D)]^{-1} k(X_D, x_2)) \quad (16)$$

*Observation error ($\varepsilon_{\text{obs},k}$)* In practice, only the noisy observation of the state $\hat{x}_k$ are available to the algorithm. Thus, the observation error quantifies the error we introduce when applying the GP to the noisy state state observation rather than the true state. This error can be characterized by assuming a small observation noise $v_k$ and linearizing the GP estimation:

$$\varepsilon_{\text{obs},k} = \hat{F}_D(x_k) - \hat{F}_D(\hat{x}_k) \\ = \hat{F}_D(\hat{x}_k - v_k) - \hat{F}_D(\hat{x}_k) \quad (17) \\ \approx \hat{F}_D(\hat{x}_k) - J_k v_k - \hat{F}_D(\hat{x}_k) = -J_k v_k,$$

where $J_k$ is the Jacobian of the GP obtained at $x_k$:

$$J_k = \nabla \hat{F}_D(x)|_{x=\hat{x}_k}. \quad (18)$$

Since the estimation $\hat{F}_D(x)$ is given by the analytical expression (7), the Jacobian can be obtained through symbolic differentiation. Note that this result is an approximation, and may fail if $F_D$ changes in smaller length scales than those of the observation noise. Here, we do not consider such scenario.

*Total error ($\varepsilon_k$)*  With all components estimated, we can now obtain an expression of the total residual at each step:

$$\begin{aligned}\varepsilon_k &= \varepsilon_{n,k} + \varepsilon_{F,k} + \varepsilon_{\text{GP},k} + \varepsilon_{\text{obs},k} \\ &= (v_{k+1} - v_k + w_k) + \varepsilon_F(x_k) - J_k v_k + \varepsilon_{D,k} \\ &= -(J_k + I)v_k + v_{k+1} + w_k + \varepsilon_F(x_k) + \varepsilon_{\text{GP},k}\end{aligned} \quad (19)$$

Note that the error is the sum of 3 elements that follow a Gaussian distribution plus a deterministic error $\varepsilon_F(x_k)$, so the complete residual follows a Gaussian distribution as well, with mean $\bar{\varepsilon}$ and covariance $\Sigma_T$:

$$\varepsilon \sim \mathcal{N}(\bar{\varepsilon}, \Sigma_T). \quad (20)$$

The mean of the residual at each step is given by

$$\begin{aligned}\bar{\varepsilon} &= \mathbb{E}[\varepsilon_k] \\ &= \mathbb{E}\big[-(J_k + I)v_k + v_{k+1} + w_k + \varepsilon_F(x_k) + \varepsilon_{\text{GP, k}}\big] \\ &= \varepsilon_F(x_k)\end{aligned} \quad (21)$$

as all the other components follow Gaussian distributions with zero mean.

Regarding the covariance, we can compute it for every pair of steps $k$ and $l$:

$$\begin{aligned}\Sigma_{kl} &= \mathbb{E}\big[(\varepsilon_k - \mathbb{E}[\varepsilon_k])(\varepsilon_l - \mathbb{E}[\varepsilon_l])^T\big] \\ &= \big[-(J_k + I)v_k - v_{k+1} + w_k + \varepsilon_{\text{GP},k}\big] \\ &\quad \times \big[-v_l^T(J_l + I)^T - v_{l+1}^T + w_l^T + \varepsilon_{D,l}^T\big] \\ &= \delta_{kl}\big[(J_k + I)\Sigma_v(J_l^T + I) + \Sigma_v + \Sigma_w\big] \\ &\quad - \delta_{k+1,l}[\Sigma_v(J_l + I)] - \delta_{k,l+1}(J_k + I)\Sigma_v + \Sigma_{\text{GP},kl},\end{aligned} \quad (22)$$

where most cross-products are 0 due to the independence in the process and observation noises. The error in the estimation of the GP depends only on the noise realization in $D$, which makes it uncorrelated with the noise in the analyzed trajectory.

The complete covariance function of the residual is then $\mathbb{E}\big[(\varepsilon - \mathbb{E}[\varepsilon])(\varepsilon - \mathbb{E}[\varepsilon])^T\big] = \Sigma_T$, where

$$\Sigma_T = \begin{bmatrix} \Sigma_{1,1} & \Sigma_{1,2} & \cdots & \Sigma_{1,n_{\text{steps}}-1} \\ \Sigma_{2,1} & \Sigma_{2,2} & \cdots & \Sigma_{2,n_{\text{steps}}-1} \\ \vdots & \vdots & \ddots & \vdots \\ \Sigma_{n_{\text{steps}}-1,1} & \Sigma_{n_{\text{steps}}-1,2} & \cdots & \Sigma_{n_{\text{steps}}-1,n_{\text{steps}}-1} \end{bmatrix}. \quad (23)$$

The covariance matrix can be used to produced a normalized residual $\tilde{\varepsilon} = \Sigma_T^{-\frac{1}{2}}\varepsilon$, with the following distribution:

$$\tilde{\varepsilon} \sim \mathcal{N}\Big(\Sigma_T^{-1/2}\big[F_Q(x) - F_D(x)\big], I\Big) \quad (24)$$

*3.5 Anomaly Detection*

The purpose of the residual is to use it to detetermine whether $Q$ was generated by the transition function $F_D$. Based on the $\chi^2$-test, we define the $p$-value of the trajectory as

$$p(\tilde{\varepsilon}) = \mathbb{P}\Big(x > \|\tilde{\varepsilon}\|^2 \,\Big|\, x \sim \chi^2_{n_x \cdot (n_{\text{steps}}-1)}\Big), \quad (25)$$

which corresponds to the probability that a nominal trajectory from $D$ produces a residual with a norm higher than $\|\tilde{\varepsilon}\|^2$. This produces a natural way of setting the threshold $p_{\text{thr}}$, by selecting the highest acceptable false positive rate, that is, probabilty of a nominal trajectory to be classified as anomalous. Our condition for classification becomes:

$$\begin{aligned}&Q \text{ is declared to not have been} \\ &\text{generated by } F_D \text{ if } p(\tilde{\varepsilon}) < p_{\text{thr}}.\end{aligned} \quad (26)$$

Finally, the complete algorithm can be summarized in these nine steps, split in offline training and online deployment:

1. **Offline training**
   1. Obtain $X_D$ and $Y_D$ from $D$.
   2. Train a GP using $X_D$ and $Y_D$.
   3. Select $p_{\text{thr}}$ according to our false positive tolerance.
2. **Online deployment**
   4. Compute observed dynamics of the new trajectory $Q$ as $\Delta\hat{x}_k = \hat{x}_{k+1} - \hat{x}_k$.
   5. Obtain residual $\varepsilon = \Delta\hat{x}_k - \hat{F}_D(\hat{x}_k)$.
   6. Compute covariance $\Sigma_T$.
   7. Normalize residual as $\bar{\varepsilon} = \Sigma_T^{-1/2}\varepsilon$.
   8. Compute $p$ from $\bar{\varepsilon}$ and $\chi^2_{n_x \cdot (n_{\text{steps}}-1)}$.
   9. Declare that $Q$ does not follow $F_D$ if $p < p_{\text{thr}}$.

4. NUMERICAL EXPERIMENTS

In this section, we present the results of two experiments that were run to assess the performance of the proposed algorithm. We evaluate its detection capability by testing many different realization of datasets $D$ and trajectories $Q$, and computing how many of them are correctly classified as anomalous depending on the $p_{\text{thr}}$ selected.

*4.1 Experimental setup*

For each experiment, we follow the following procedure:

1. We generate 500 different datasets $D$ using a nominal system $f_D$, each one with $N_D$ trajectories with $N_\ell$ points, using different realizations of $w_k$.
2. We train 20 GPs using 20 of those datasets and optimizing the hyperparameters.
3. We train 480 additional GPs with the remaining datasets, using as hyperparameters the average of those obtained in the first 20 trainings.
4. We generate 800 trajectories using $f_Q$ with random realizations of $w_k$ and $v_k$.
5. For each pair of trajectory and GP, we obtain a $p$-value using the algorithm described in Section 3. The complete collection of values of $p$ is used determine the ratio of trajectories with $p$-value below $p_{\text{thr}}$, for a range of thresholds between 0 and 1.

By generating many datasets and trajectories, we can obtain numerically a reliable approximation of the actual distribution of $\tilde{p}$ when the algorithm is run on a trajectory generated by $f_Q$. A repository with the implementation of the algorithm and the experiments can be found in https://github.com/AlejandroPenacho/GP-system-discrimination.

## 4.2 Damped pendulum with mass imbalance

We define a nominal and an anomalous systems given by

$$f_D(x) = \begin{bmatrix} x_2 \\ -\sin(x_1) - 0.1x_2 \end{bmatrix},$$
$$f_Q(x) = \begin{bmatrix} x_2 \\ -\sin(x_1 + 0.3) - 0.1x_2 \end{bmatrix}, \quad (27)$$

with sampling time $\Delta T = 0.3$.

To visualize the system, we show in Fig. 1 the trajectories in the training dataset $D$, together with 5 realizations of a trajectory generated with $f_Q$. The algorithm, then, must determine if these trajectories has been produced by the same system that produced the dataset $D$. The training dataset has 10 trajectories of 14 steps each, generated with $\Sigma_w = 10^{-2}I$, $\Sigma_v = 0$, starting from randomly selected initial positions in $[-2, 2] \times [-2, 2]$. The trajectories generated with $f_Q(x)$ use $\Sigma_w = \Sigma_v = 4 \cdot 10^{-4}I$ and start from $x_0 = [0.7 \; 0.4]$.

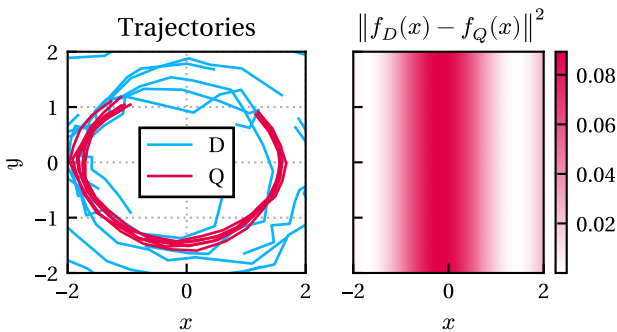

Fig. 1. Setup for the pendulum example. Left plot: trajectories in $D$ with 5 realizations of a trajectory generated with $f_Q$. Right plot: squared error between the dynamics of $f_Q$ and $f_D$.

We follow the procedure explained in Section 4.1 to obtain a total of 400,000 $p$-values, and we analyze their distribution in Fig. 2, left plot. Since $p_{\text{thr}}$ corresponds to the ratio of trajectories generated by $f_D$ that are classified as anomalous, this plot is equivalent to the Receiver Operating Characteristic curve commonly used in binary testing.

In the plot we can observe that, setting a threshold of $p_{\text{thr}} = 0.2$, we get a detection probability of 60% for trajectories generated with $f_Q$ when we analyze the first 10 steps of the trajectory. Interestingly, this detection decreases to around 45% when use 20 steps, implying that analyzing less steps leads to better results.

To understand the reason, we show in the right plot of Fig. 2 the covariance of $\varepsilon$ at each step, together with the square error produced by the difference between $F_D(x)$ and $F_Q(x)$. This difference becomes noticeable only between steps 8 and 11, and is what allows the algorithm to detect a difference in the dynamics. Between steps 11 and 20, the difference in the dynamics is minimal compared to the covariance of the residual, so analyzing 20 steps instead 10 does not provide an improvement in the detection.

For the right plot of Fig. 2 we can also observe the uncertainty produced by the GP approximation (green area) and that produced by the noise (blue area). This results provide an important insight on how the detection can be improved: if we collect more data, we can reduce the uncertainty $\sigma_{\text{GP}}$. Meanwhile, we would need to decrease the observation noise $v$ if we want to reduce $\sigma_{\text{n}}$. In this experiment, both have a similar contribution to the total uncertainty, so both options are equally valid to improve the effectiveness of the algorithm.

## 4.3 Van der Pol oscillator

We now consider a second system with nominal and anomalous dynamics given by

$$f_D(x) = \begin{bmatrix} x_2 \\ 0.4(1 - x_1^2)x_2 - x_1 \end{bmatrix},$$
$$f_Q(x) = \begin{bmatrix} x_2 \\ 0.55(1 - x_1^2)x_2 - x_1 \end{bmatrix}, \quad (28)$$

again with sampling time $\Delta T = 0.3$. As with the case of the pendulum, we show in Fig. 3 the trajectories used to train the GP and 5 realizations of the trajectory generated by $f_Q(x)$. For this problem, the training set has 30 trajectories of 8 steps each, generated with $\Sigma_w = 0.01I$ and $\Sigma_v = 0$. The trajectories $Q$ have $\Sigma_w = \Sigma_v = 10^{-3}$, and start at $x_0 = [0.7 \; 0.4]$.

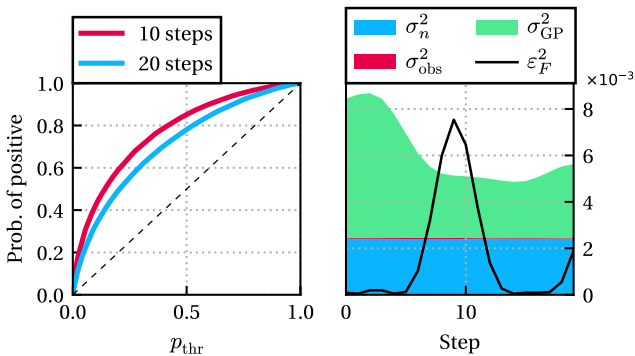

Fig. 2. Results for the pendulum example. Left plot: ROC curve obtained using different number of steps for the classification. Right plot: covariance and square error at each step of the analyzed trajectories.

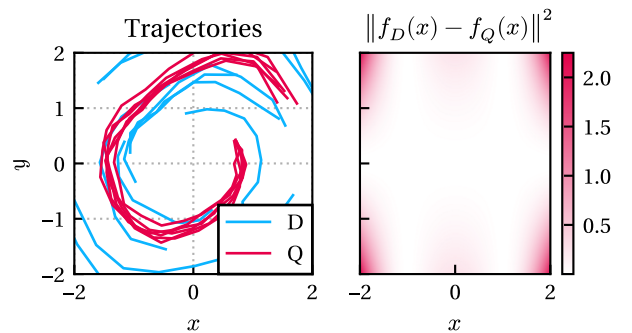

Fig. 3. Setup for the Van der Pol example. Left plot: trajectories in $D$ and 5 realizations of a trajectory generated with $f_Q$. Right plot: squared error between the dynamics of $f_Q(x)$ and $f_D(x)$.

Following again the procedure from Section 4.1, we obtain the probability of classifying a trajectory generated with $f_Q$ as anomalous as a function of the threshold $p_{\text{thr}}$ and the number of steps we consider.

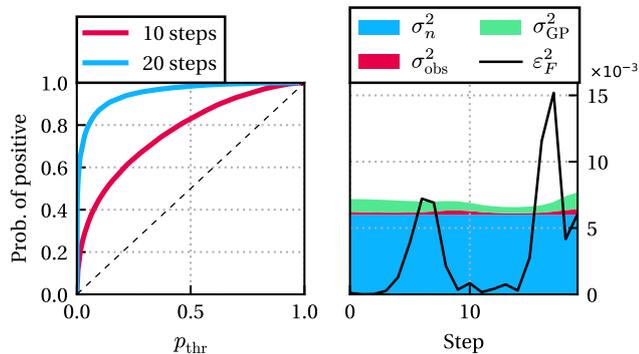

Fig. 4. Results for the Van der Pol example. Left plot: ROC curve obtained using different number of steps for the classification. Right plot: covariance and square error at each step of the analyzed trajectories.

For this problem, the results are significantly better than for the damped pendulum: analyzing 20 steps, we can achieve a detection probability of 80% with $p_{\text{thr}} = 0.05$. Also, we can now see how the results obtained analyzing 20 steps are better than analyzing 10 steps. To understand the difference with the previous example, we look at the squared residual produced by the difference in the dynamics compared with the covariance of the total residual in Fig. 4, right plot. We see that between steps 15 and 18, the value of $\varepsilon_F^2$ increases well above the covariance of the residual. By taking residuals in these steps, the detection probability increases significantly.

If we compare the uncertainty produced by the noise ($\sigma_n^2$) and by the GP ($\sigma_{\text{GP}}^2$), we can observe that the second is much smaller, mostly due to the fact that the training dataset has more trajectories than in the pendulum example. This difference also indicates that the function approximation has a high level of confidence, and having a larger dataset $D$ would then not improve the detection capability of the algorithm. The bottleneck for the detection, in this case, is in the covariance of the noise $\sigma_n^2$, which can be improved by reducing the observation noise.

## 5. CONCLUSION

We have established a model-free approach to discrimination analysis for dynamical systems, used to test whether a trajectory fits with the dynamics observed in a dataset. In this approach, we use a GP to approximate the dynamics of the dataset, obtaining a measure of the uncertainty of the approximation. This uncertainty is combined with the effect of the process and observation noise to obtain a distribution of the residual between the observed and predicted dynamics, so we can apply the concept of $p$-value to the new trajectory.

This first work into establishing the framework hightlihgts several research directions to explore in the future. First of all, the capability to quantify the uncertainty of the estimated dynamics can be used to design optimal sampling techniques to maximize the detection capability of the algorithm. Second, we can use prior knowledge of the nominal dynamics of the system to improve the approximation of the GP, possibly by modifying its kernel function. Last, we can use knowledge of the possible deviation of the dynamics of the system to, for example, weight the residual obtained in certain states higher than others, in order to fine-tune the algorithm to different anomalies. In any case, by retaining the uncertainty quantification properties of the algorithm, we can maintain the statistical properties of the algorithm that let us select a desired false positive rate with $p_{\text{thr}}$.